\documentstyle[manuscript,eqsecnum,aps,epsfig]{revtex}
\def\ep {\epsilon}
\begin{document}
\preprint{\begin{tabular}{l}
\hbox to\hsize{\mbox{ }\hfill SNUTP 00-020}  \\ [-3mm]
\hbox to\hsize{\mbox{ }\hfill KIAS-P00052}  \\ [-3mm]
\end{tabular}}

\title{Massive field contributions to the QCD vacuum \\tunneling amplitude}
 
\author{O-Kab Kwon$^{a}$\footnote{Electronic address; kok@phya.snu.ac.kr}
, Choonkyu Lee$^{a,b}$\footnote{Electronic address; cklee@phya.snu.ac.kr},
and Hyunsoo Min$^{c}$\footnote{Electronic address; hsmin@dirac.uos.ac.kr}}

\address{$^{a}$ Physics Department and Center for Theoretical Physics,
Seoul National University, Seoul \\[-2mm]151-742, Korea}
\address{$^{b}$ School of Physics, Korea Institute for Advanced Study,
207-43 Cheongryangryi-Dong,\\[-2mm] Dongdaemun-Gu, Seoul 130-012, Korea}
\address{$^{c}$  Physics Department, University of Seoul, 
Seoul 130-743, Korea}
\maketitle
\vskip -0.7cm 
\begin{abstract}
For the one-loop contribution to the QCD vacuum tunneling amplitude
by quarks of generic mass value, we make use of a calculational scheme 
exploiting a large mass expansion together with a small mass expansion. The 
large mass expansion for the effective action is given by a series 
involving higher-order Seeley-DeWitt coefficients, and we carry this 
expansion up to order $1/(m\rho)^8$, where $m$ denotes mass of the quark 
and $\rho$ the instanton size parameter. For the small 
mass expansion, we use the known exact expression for the particle 
propagation functions in an instanton background and evaluate explicitly 
the effective action to order $(m\rho)^2$. A smooth interpolation of  
the results from both 
expansions suggests that the quark contribution to the 
instanton tunneling amplitude have a relatively simple $m\rho$-dependent
behavior. 
\end{abstract}

\section{Introduction}
Instantons\cite{belavin,jackiw}, as localized finite-action solutions of the 
Euclidean Yang-Mills field equations, describe vacuum tunneling and are
believed to have important nonperturbative roles in low energy QCD. For 
an excellent review on instantons in QCD and general gauge theories, see 
Refs.\cite{schaffer,shifman}. For actual instanton calculations, one needs 
to know above all the one-loop tunneling amplitude or the Euclidean 
one-loop effective action in the background field of a single 
(anti-)instanton. The latter quantity is thus of fundamental importance 
in instanton physics, and in the zero mass limit of scalar or quark fields 
'tHooft\cite{thooft} was able to calculate the appropriate one-loop 
contribution exactly. But, with finite quark mass, such exact calculation 
does not look feasible and one has to be satisfied with approximate 
results. [Note that, aside from up and down quarks, all other quarks possess 
sizable mass]. In this paper we shall describe our approach to determine 
the quark mass dependence in the one-loop vacuum tunneling amplitude, 
and report some new results from this analysis.

 By studying the field-theoretic effective action one can take 
systematically quantum nature of the fields into account, and already 
at the one-loop level it has provided us with certain relevant 
information on various physically significant effects\cite{schwinger}. In 
particular, the leading-order renormalization group coefficients in 
field theories are encoded in the divergences of the corresponding 
bare one-loop effective action. These divergent terms can be found most 
simply with the help of the background field method\cite{dewitt,thooft75} 
and the Schwinger-DeWitt proper-time algorithm\cite{schwinger,dewitt};
they are entirely given by the second Seeley-DeWitt coefficient 
$\tilde{a}_2$\cite{dewitt,seeley,lee} 
in four-dimensional space-time. (For a recent literature discussing 
this method, see Ref.\cite{hochberg}). But the evaluation of the 
full finite part of the one-loop effective action in any non-trivial 
background field corresponds to a formidable mathematical problem in 
general. For an approximate calculation (in a slowly-varying 
background) the so-called derivative expansion of the effective 
action has been utilized by various authors\cite{goldstone,chan}. 

 As for the contribution to the instanton one-loop effective action by 
spin-0 or spin-1/2 matter fields of, say, mass $m$, we shall consider 
both the approximation applicable for relatively large $m\rho$ ($\rho$ 
is the instanton size), i.e., large mass expansion and the mass 
perturbation scheme useful for relatively small $m\rho$. Note 
that the nature of the approximation is governed by the dimensionless 
parameter $m\rho$.(Dependence on the renormalization mass scale 
$\mu$ can be treated separately). The large mass expansion is essentially 
a series involving higher-order Seeley-DeWitt coefficients, for which 
a simple computer algorithm has been developed 
recently\cite{belkov,fliegner,van}.
We 
then make a smooth interpolation of the results found in those two 
different regimes, with the expectation that some general 
pattern, which is meaningful over a wide range of mass values, 
may emerge. This information should be valuable in phenomenological 
studies related to instanton effects. To connect the amplitude given for  
different mass scales, one should be careful about possible large 
finite-renormalization effects and renormalization schemes used. In 
this paper we treat various issues 
related to this general idea in a reasonably self-contained manner. 

 In Sec.{\rm II} we present a concise review on the Schwinger proper-time 
representation of the effective action, various renormalization schemes, and 
the large mass expansion. Also discussed are finite renormalization 
effects specific to renormalization prescriptions chosen, 
since they can introduce additional mass (as well as renormalization 
scale) dependences into the effective action. This understanding will 
become important when one has to change the results obtained in 
one renormalization prescription to that in another prescription.
 
 In Sec.{\rm III} the one-loop effective action for a massive scalar field 
in a constant Yang-Mills field background is considered to see how our 
general scheme would fit in for this simple case. Here we make a 
detailed comparison between the known, exact, effective action (given 
in a single integral form) and the corresponding result based on the 
large mass expansion. 
 
 In Sec.{\rm IV} the spin-0 one-loop effective action in a Yang-Mills 
instanton background is studied on the basis of the large mass 
expansion. (Contributions due to fields of different spin can be 
related to this spin-0 amplitude). We consider up to the sixth 
Seeley-DeWitt coefficient. Here 
our finding is that, for $m\rho\agt 1.8$, the large mass expansion 
appears to give a good approximation to the effective action. 

 In Sec.{\rm V}  we study the spin-0 instanton effective action for 
small $m\rho$, utilizing the known expressions for the massless 
propagators\cite{brown} in an instanton background and the mass  
perturbation. Since the naive mass perturbation leads to a logarithmically 
divergent integral\cite{carlitz}, a suitably modified perturbation method must 
be employed to obtain a well-defined small-mass correction term. We here 
reconfirm the ${\cal O}((m\rho)^2\ln m\rho)$ term previously found in 
Ref.\cite{carlitz}, and provide for the first time the full 
${\cal O}((m\rho)^2)$ contribution to the instanton 
effective action. 

 In Sec.{\rm VI}  we consider an interpolation of our amplitude to intermediate 
values of $m\rho$, given the results of the previous two sections. Here 
we also make appropriate changes in our results so that they may describe 
the spin-1/2 instanton effective 
action; this result is directly relevant for quarks with 
nonzero mass. Note that, 
due to the hidden supersymmetry in an instanton background, one can 
utilize the result for the spin-0 case to find the contribution 
due to spin-1/2 fields\cite{thooft}. 

 In Sec.{\rm VII} we conclude with some remarks. In Appendix A some explicit 
expressions for higher-order Seeley-DeWitt coefficients can be 
found. Appendix B contains an analysis of a certain function which 
figures in our small-mass expansion of Sec.{\rm V}.

\section{The one-loop effective action, 
renormalization, and the large mass expansion 
~~~~~~~~~~~~~~~~~~~~~~~~~~~~~~~~~~~~~~~~~~~~~~~~~~~~~~~~~~~~}

To be definite, we will consider a four-dimensional, Euclidean, 
Yang-Mills theory with matter described by {\it complex~} scalar or 
Dirac spinor 
fields of mass $m$. Then, in any given Yang-Mills background fields 
$A_\mu^a(x)$, one may represent the (Pauli-Villars regularized) one-loop 
effective action due to matter fields by
\begin{equation}
\Gamma(A) = \lambda \ln \left[
{{\rm Det}(G^{-1} + m^2) \over {\rm Det} (G^{-1}_0+m^2)}
{{\rm Det}(G^{-1}_0 + \Lambda^2) \over {\rm Det} (G^{-1}+\Lambda^2)}
\right].
\end{equation}
Here, $\lambda = 1(-{1\over 2})$ for scalar(spinor) fields, $\Lambda$ is the 
large regulator mass, $G^{-1}$ stands for the appropriate quadratic differential
operator, viz., 
\begin{equation}
G^{-1} =  
\left\{ \begin{array} {cl}  -D^2, &\qquad \mbox{(for scalar)} \\
            (\gamma D)^2, &\qquad \mbox{(for spinor)}, 
\end{array}\right.
\end{equation}
and $G_0^{-1} = G^{-1}|_{A_\mu=0}=-\partial^2$. [Also, $D^2=D_\mu D_\mu$ 
and $\gamma D = \gamma_\mu D_\mu$ with the 
covariant derivative $D_\mu=\partial_\mu -i A^a_\mu T^a 
\equiv \partial_\mu -i A_\mu$ ( $T^a$ denote the group generators in the 
matter
representation satisfying the commutation relations
$[T^a,T^b]=i f_{abc}T^c$ ), 
and our $\gamma$-matrices, which are antihermitian, satisfy the 
relations $\{\gamma_\mu, \gamma_\nu\} = -2 \delta_{\mu\nu}$].

In the proper-time representation\cite{schwinger,dewitt}, one represents
 $\Gamma(A)$ by
\begin{equation}
\Gamma(A) = -\lambda \int^\infty_0 { ds \over s} 
(e^{-m^2 s} - e^{-\Lambda^2 s} ){\rm Tr} [e^{-s G^{-1}}- e^{-s G_0^{-1}} ],
\end{equation}
where `Tr' denotes the trace over space-time coordinates and all other discrete
indices. More explicitly, writing  ${\rm Tr}=\int d^4x \, \rm {tr}$
and introducing the proper-time Green function  
$\Big <x s\Big |y\Big >$ $ = \Big\langle x \Big | e^{-s G^{-1}}\Big |y\Big > $,~  
$\Gamma(A)$ can be expressed as
\begin{equation}
\Gamma(A) =-\lambda \int^\infty_0 { ds \over s} 
(e^{-m^2 s} - e^{-\Lambda^2 s} ) \int d^4x \, {\rm tr}
 \left[\Big\langle xs\Big |x\Big > - \Big\langle xs\Big |x\Big >|_{A_\mu=0} 
\right], \label{propertime}
\end{equation}
The full effective action is thus determined if the coincidence limit (i.e., 
$y=x$) of the proper-time Green function is known. The 
expression (\ref{propertime}) diverges
logarithmically as we let $\Lambda \to \infty$; to isolate such 
divergent pieces, we may exploit the asymptotic expansion\cite{dewitt,seeley}
\begin{equation}\label{series}
s\to0+: \qquad \Big\langle xs\Big |y\Big >={1 \over (4\pi s)^2} 
e^{-{(x-y)^2 \over 4s}}
\left\{ \sum^\infty_{n=0} s^n a_n(x,y)\right\},
\end{equation}
where the leading coefficient has the coincidence limit $a_0(x,x)=1$.
Using this expansion in (\ref{propertime}), we then see that the divergences
in $\Gamma(A)$ as $\Lambda\to\infty$ are related to the coincidence limits
$\tilde a_1(x)\equiv {\rm tr} a_1(x,x)$ and 
$\tilde a_2(x)\equiv {\rm tr} a_2(x,x)$,
which correspond to the first and second Seeley-DeWitt coefficients respectively.
Simple calculations yield
\begin{eqnarray}
\tilde a_1(x)&=&0, \qquad \qquad  \mbox{(for both scalar and spinor)} \\
\tilde a_2(x)&=& \left\{ \begin{array}{cl} -{1\over12}\, {\rm tr}
 ( F_{\mu\nu}(x) F_{\mu\nu}(x)),& \qquad\mbox{(for scalar)} \\
{2\over3} \, {\rm tr}
 ( F_{\mu\nu}(x) F_{\mu\nu}(x)),& \qquad\mbox{(for spinor)},
\end{array}\right.
\end{eqnarray}
where $F_{\mu\nu}\equiv F_{\mu\nu}^a T^a=i[D_\mu,D_\nu]$.  
(The `tr' here refers to the trace over gauge group representation indices
only). Based on these, we may now write the above effective action for
$m^2\ne0$ as 
\begin{equation}\label{gamma}
\Gamma(A)=Y(\ln{\Lambda^2 \over m^2}) \int
d^4x F^a_{\mu\nu} F^a_{\mu\nu}  + \overline{\Gamma}(A)
\end{equation} with
\begin{equation}
Y=\left\{ \begin{array}{cl}{1\over12} {C\over(4\pi)^2},
& \qquad\mbox{(for scalar)} \\
{1\over3}{C\over(4\pi)^2},& \qquad\mbox{(for spinor), }
\end{array}\right.
\end{equation}
($C$ is defined by ${\rm tr} (T^a T^b)=\delta_{ab} C$), and then the contribution
\begin{equation}\label{gammabar}
\overline{\Gamma}(A)=-\lambda 
\int^\infty_0 {ds \over s^3} e^{-m^2s}\int d^4x 
\left[1-(1+s{\partial \over \partial s}+ {1\over2} 
s^2{\partial^2 \over \partial s^2})|_{s=0}\right]
{\rm tr} \left(s^2 \Big\langle xs\Big |x\Big >\right)
\end{equation}
becomes well-defined as long as $m^2$ is non-zero. [In (\ref{gammabar}),
$(1+s{\partial \over \partial s}+ {1\over2} 
s^2{\partial^2 \over \partial s^2})|_{s=0} f(s)\equiv
f(0)+sf'(0) + {1\over2} s^2 f''(0)$].

The logarithmic divergence in $\Gamma(A)$ is canceled by the 
renormalization counterterm associated with the coupling constant
renormalization of the classical (bare) action ${1\over4g_0^2} \int d^4x
F^a_{\mu\nu}F^a_{\mu\nu} $. But the resulting renormalized
one-loop amplitude 
depends on the renormalization prescription chosen. 
From the very structure exhibited in (\ref{gamma}), our amplitude
$\overline{\Gamma}(A)$ can be considered as defining {\em a} 
renormalized one-loop effective action; but, this prescription
cannot be used for the strictly massless case.
Instead, one may here consider adding 
to $\Gamma(A)$ the counterterm
\begin{equation}
\Delta \Gamma(A)=-Y(\ln{\Lambda^2\over \mu^2}) \int d^4x   
F^a_{\mu\nu}F^a_{\mu\nu}
\end{equation}
($\mu$ is an arbitrarily introduced renormalization mass) to obtain
the renormalized one-loop effective action
\begin{equation}\label{gammaren}
\Gamma_{\rm ren}(A)=-Y(\ln{m^2\over \mu^2}) \int d^4x   
F^a_{\mu\nu}F^a_{\mu\nu} + \overline{\Gamma}(A),
\end{equation}
where $\overline{\Gamma}(A)$ is defined by (\ref{gammabar}). It
should be remarked that $\Gamma_{{\rm ren}}(A)$, given by (\ref{gammaren}),
is expected to have a well-defined limit for $m^2 \to 0$ (i.e., does not 
exhibit infrared singularities), if the operator $G^{-1}$ does not allow
any normalizable zero eigenmode.

Other renormalization prescriptions may also be chosen. Let 
$\Gamma_{\rm MS}(A)$ denote the renormalized amplitude in the so-called 
minimal subtraction scheme\cite{thooft72} associated with the dimensional 
regularization. Then, to obtain the expression for 
$\Gamma_{\rm MS}(A)$  in the scalar case, one should add to that of 
$\Gamma_{\rm ren}(A)$ the following finite renormalization term
\begin{equation}
 {C\over(4\pi)^2 \cdot 12} (\ln4\pi -\gamma)
\int d^4x  F^a_{\mu\nu}F^a_{\mu\nu},
\end{equation}
where $\gamma=0.5772\cdots$ is the Euler's constant. 
In the thus found expression of $\Gamma_{\rm MS}(A)$ the mass parameter 
$\mu$, which enters the expression through 
 $\Gamma_{\rm ren}(A)$, describes the normalization mass conventionally
introduced in the minimal subtraction scheme. As for the expression of 
$\Gamma_{\rm MS}(A)$ in the spinor case, the finite renormalization
term to be added to that of $\Gamma_{\rm ren}(A)$
turns out to be 
\begin{equation}\label{spinconnect}
 {C\over(4\pi)^2 \cdot 3} (\ln4\pi -\gamma)
\int d^4x  F^a_{\mu\nu}F^a_{\mu\nu}.
\end{equation}
[This is the case when the spinor trace of 1 is taken to be four].
In another often-used prescription, one specifies the renormalization 
counterterm via the momentum-space subtraction scheme, i.e., by imposing 
a normalization condition at certain external momentum
value, $p^2=\mu^2$. Then the corresponding renormalized expression,
$\Gamma_{\rm mom}(A)$, is given by that of  $\Gamma_{\rm ren}(A)$
plus  the finite renormalization term

\begin{eqnarray}
 {C\over(4\pi)^2 \cdot 12}
&&\left[
-{8\over3} + \ln{m^2\over \mu^2} - 8 {m^2\over \mu^2}
    +\left(1+4{m^2\over \mu^2}\right)^{3\over2}
    \ln\left( {\sqrt{1+4{m^2\over \mu^2}} +1} \over
     \sqrt{1+4{m^2\over \mu^2}} -1\right) 
 \right]   \\ 
&&\times\int d^4x  F^a_{\mu\nu}F^a_{\mu\nu} , 
 \hspace{9 cm} \mbox{(for scalar)}\nonumber 
\end{eqnarray}
or
\begin{eqnarray}
 {C\over(4\pi)^2 \cdot 3}
&&\left[
	-{5\over3} +\ln{m^2\over \mu^2} + 4 {m^2\over \mu^2}
    +\left(1-2{m^2\over \mu^2} \right)\sqrt{1+4{m^2\over \mu^2}}
    \ln\left( {\sqrt{1+4{m^2\over \mu^2}} +1 \over
     \sqrt{1+4{m^2\over \mu^2}} - 1 }\right) 
 \right] \\
&&\times\int d^4x  F^a_{\mu\nu}F^a_{\mu\nu} 
, 
\hspace{9 cm} \mbox{(for spinor)} \nonumber
\end{eqnarray}
with the corresponding reinterpretation of the parameter $\mu$. These
renormalization-prescription dependences of the one-loop effective action are 
of course explained by the fact that the tree-level contribution involves,
as a multiplicative factor, the renormalized coupling ${1\over g_R^2}$ (whose
value may vary with renormalization prescriptions).

Thanks to the exact connection formulas we have described above, knowledge on
the one-loop effective action in one renormalization 
prescription can immediately be changed into that in another prescription. In 
fact, in theories containing several matter fields of different mass 
scales ( e.g., QCD with quarks of very different masses), one 
may well adopt 
different renormalization prescriptions for different matter field loops. We 
here note that use of the minimal subtraction for a heavy-quark loop is 
rather unnatural, due to the lack of manifest decoupling\cite{appelquist}. But 
this is not an issue in our discussions.

 The next task is to find the actual full expression for the one-loop
effective action --- at present, this is possible only with a background 
field of very special character. But, if the mass parameter is sufficiently
large, it can be studied for generic smooth background fields by utilizing a 
systematic large-mass expansion, which is obtained by inserting the 
asymptotic expansion (\ref{series}), say, into the formula (\ref{gammabar}) for 
$\overline{\Gamma}(A)$. This assumes the form
\begin{equation} \label{gammabarmass}
\overline{\Gamma}(A)= - {\lambda\over (4\pi)^2}  \sum_{n=3}^\infty
{(n-3)! \over (m^2)^{n-2} }\int d^4 x \tilde{a}_n(x), \hskip 1cm 
(\tilde{a}_n(x)\equiv {\rm tr} a_n(x,x) ).
\end{equation}
That is, for large enough mass, we have the one-loop effective action 
(in any renormalization prescription) expressed by a series 
involving higher-order 
Seeley-DeWitt coefficients $\tilde{a}_n(x)$ ($n\ge 3$), the calculation
of which may be performed using a computer \cite{belkov,fliegner,van}. 
If only the 
leading term is kept with the 
series (\ref{gammabarmass}), one find, explicitly,
\begin{equation}\label{gammabarthree}
m\rightarrow \infty:\quad
\overline{\Gamma}(A) =
\left\{
\begin{array}{cl} -{1\over16\pi^2}{1\over m^2} {1\over180}
\int d^4x \ {\rm tr} [{3\over2}(D_\mu F_{\nu\lambda}) (D_{\mu}F_{\nu\lambda})
-4i
F_{\mu\nu}F_{\nu\lambda}F_{\lambda\mu}],
& \qquad \mbox{(for scalar)} \\
 {1\over32\pi^2}{1\over m^2} {2\over45}
\int d^4x \ {\rm tr} [-3(D_\mu F_{\nu\lambda}) (D_\mu F_{\nu\lambda})
+13i F_{\mu\nu}F_{\nu\lambda}F_{\lambda\mu}],
& \qquad \mbox{(for spinor)}
\end{array} \right.
\end{equation}
where $D_\mu F_{\nu\lambda}\equiv [D_\mu,F_{\nu\lambda}]$. If the 
background fields
under consideration satisfy
the classical Yang-Mills field equations, one can show using the Bianchi
identities and the property of trace that $\int d^4x \  {\rm tr} 
[(D_\mu F_{\nu\lambda}) (D_{\mu}F_{\nu\lambda})] = 4i \int d^4x \ {\rm tr} 
[F_{\mu\nu}F_{\nu\lambda}F_{\lambda\mu}] $. Hence, for the
{\it on-shell} effective action, (\ref{gammabarthree}) can be further simplified
as 
\begin{equation}
m\rightarrow \infty:\quad
\overline{\Gamma}(A) =
\left\{
\begin{array}{cl} -{1\over16\pi^2}{1\over m^2} {i\over 90}
\int d^4x \ {\rm tr} (F_{\mu\nu}F_{\nu\lambda}F_{\lambda\mu}),
& \qquad \mbox{(for scalar)} \\
 {1\over16\pi^2}{1\over m^2} {i\over45}
\int d^4x \ {\rm tr} (F_{\mu\nu}F_{\nu\lambda}F_{\lambda\mu}),
& \qquad \mbox{(for spinor)}.
\end{array} \right.
\end{equation}
For some explicit expressions of the higher-order Seeley-DeWitt
coefficients, see Sec.{\rm IV} and Appendix A. Also note 
that the large mass expansion for
the effective action in other renormalization schemes can be
obtained from the expansion (\ref{gammabarmass}) for 
$\overline{\Gamma}(A)$ and the exact connection formulas.

 The large-mass expansion is only an asymptotic series, and 
the useful range of the series (\ref{gammabarmass}) (as regards 
the magnitude of $m$) will depend much on the nature of the 
background field and also on some characteristic scale(s) 
entering the background. For a sufficiently smooth background, this 
large-mass expansion may be used to obtain a reliable approximation 
to the effective action even for moderately large values of $m$. But 
the series (\ref{gammabarmass}) is bound to lose the predictive power 
for `small' values of $m$, and for the small-$m$ effective action 
one should employ a totally different strategy, such as the small-mass 
expansion if its exact expression in the massless limit has been 
known by some other methods. In the next section, we shall first 
see how good the large-mass expansion can be for the much studied 
case of the one-loop effective action in the constant Yang-Mills 
field background. Also considered is its small-mass expansion which 
may serve, together with the result of the large mass expansion, as 
a basis to infer the behavior of the effective action 
for arbitrary mass.

\section{THE SPIN-0 EFFECTIVE ACTION IN A CONSTANT SELF-DUAL
 YANG-MILLS FIELD BACKGROUND
~~~~~~~~~~~~~~~~~~~~~~~~~~~~~~~~~~}

In this section, various approximation schemes to be used later will be 
tested against the exact result, choosing a rather simple background 
field.  In non-Abelian gauge theories, a constant field strength
is realized either by an Abelian vector potential which varies linearly
with $x^\mu$  or by a constant vector
potential whose components do not commute\cite{leutwyler}.
In this paper we only consider the
case of the Abelian vector potential.
Assuming the SU(2) gauge group, an Abelian vector potential can 
be written as $A_\mu=-{1\over4} f_{\mu\nu}x_\nu \tau^3$ (with the field
strength tensor $F_{\mu\nu}=f_{\mu\nu}\tau^3/2$), where $\tau^3$ is the 
third Pauli matrix. If we further restrict our attention to that with 
the self-dual
field strength (i.e., $F_{\mu\nu} = \frac{1}{2} \epsilon_{\mu\nu\lambda\delta}
F_{\lambda\delta}$), its nonzero components may be specified by setting
 $f_{12}=f_{34}=H$ with the constant `magnetic' field $H$.

In this Abelian constant
self-dual field, let us consider the one-loop effective action
due to an isospin-1/2, spin-0 (complex-valued) matter field, 
taking the mass $m$ of
our spin-0 field to be relatively large so that
the large mass expansion (\ref{gammabarmass}) may be used.
For this case,
some leading  Seeley-DeWitt coefficients are easily evaluated 
(using the formulas given in Appendix A, for instance),
\begin{eqnarray} \label{const_an}
\tilde{a}_2 &=& -{2\over3} (H/2)^2, \quad
\tilde{a}_4 = {2\over 15} (H/2)^4, \\  \nonumber
\tilde{a}_6 &=& -{4\over 189} (H/2)^6, \quad
\tilde{a}_8 = {2\over 675}(H/2)^8.
\end{eqnarray}
Note that we get zero for all odd coefficients here.
Using these values, we then find that, for relatively large $m$, 
the effective action is given by the series 
\begin{equation}\label{inversemass}
\overline{\Gamma}(H;m)= {VH^2 \over 16\pi^2}
\left( -{1\over 120} ({H\over m^2})^2
+ {1\over 504 } ({H\over m^2})^4
-{1\over 720} ({H\over m^2})^6
+\cdots \right),
\end{equation}
where $V$ denotes the four-dimensional Euclidean volume.

For this case, it is actually not difficult to find the  {\em exact}
expression for the one-loop
effective action, following closely  Schwinger's original analysis in
QED\cite{schwinger}.  After some algebras, one finds the trace of
the proper-time Green function to be given by\cite{yildiz}
\begin{equation} \label{constpgreen}
{\rm tr}\left< xs|x\right>=
 {2\over(4\pi s)^2}
\left[{(Hs/2)^2 \over \sinh^2(Hs/2)}\right].
\end{equation}
One can easily check that the expressions given in (\ref{const_an}) 
are correct ones by considering a small-$s$ series of this exact 
expression. Inserting (\ref{constpgreen}) into the formula (\ref{gammabar}) 
then yields the exact expression
\begin{equation} \label{constbar}
\overline{\Gamma}(H;m)=- 2V \int^\infty_0 {ds\over s}
 e^{-m^2 s} {1\over(4\pi s)^2}
\left[{(Hs/2)^2 \over \sinh^2(Hs/2)} -1+ {1\over3} (Hs/2)^2 \right].
\end{equation}
\begin{figure}
\begin{center}
\epsfxsize=4in
\epsffile{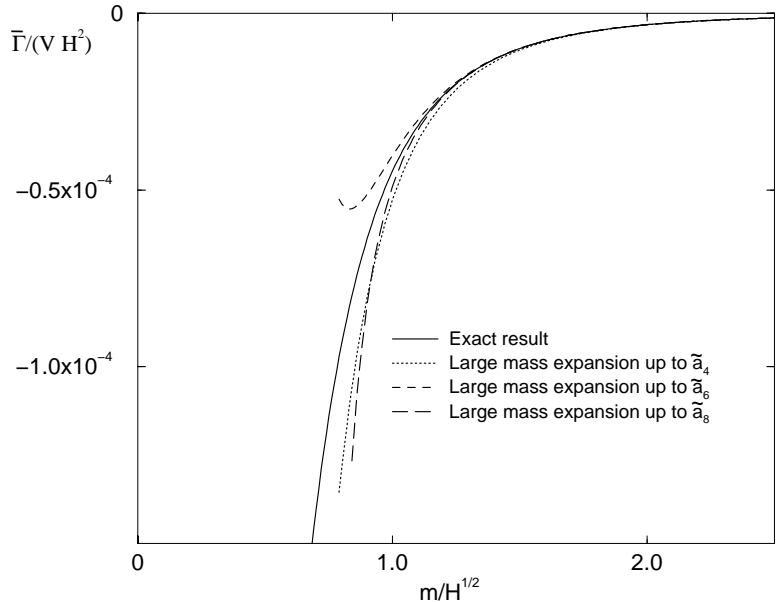}
\caption{Plot of the effective action $ \overline\Gamma(H;m)$.}
\end{center}
\end{figure}
\begin{flushleft}Comparing the result of large mass 
expansion in (\ref{inversemass})
against this exact result, we can investigate the validity
range of the former. From the plots in Fig.1, it should be
evident that for mass values in the range $m/\sqrt{H}\agt 1$,
summing only a few leading terms in the series (\ref{inversemass}) already
produces the results which are very close to the exact one. The large 
mass expansion is useful if $m \agt \sqrt{H}$.  \\
~~~~Now suppose that the exact expression (\ref{constbar}) were not
available to us. For mass value not larger than $\sqrt{H}$, 
the large mass expansion (\ref{inversemass}) fails
to give useful information. Nevertheless, if one happens to know the
one-loop effective action for {\em small} mass, this additional
information and the large mass expansion might be used to infer
the behavior of the effective action for
general, small or large, mass.[Note that, in an instanton 
background, this becomes a real issue since the full $m$-dependence 
of the effective action is not known there]. In exhibiting this,
$\overline{\Gamma}(H;m)$ will not be convenient since it becomes
ill-defined as $m\to0$. So, based on the relation (\ref{gammaren}), we may
consider the renormalized action $\Gamma_{\rm ren}(H;m,\mu)$ given 
by \end{flushleft}
\begin{equation}\label{constren}
\Gamma_{\rm ren}(H;m,\mu)=-  { V H^2 \over(4\pi )^2 \cdot 6}
\ln({m^2 \over \mu^2})
 + \overline{\Gamma}(H;m).
\end{equation}
which is well-behaved for small $m$. Large mass expansion for
$\Gamma_{\rm ren}(H;m,\mu)$ results once if the expansion (\ref{inversemass})
is substituted in the right hand side of (\ref{constren}). 

To find the small-$m$ expansion, we find it convenient to consider the 
quantity 
\begin{eqnarray}\label{gammatilde}
Q(H;m)&\equiv&
\int_0^{m^2} d\bar{m}^2 {\partial  \over \partial \bar{m}^2}
 \Gamma_{\rm ren}(H;\bar{m},\mu) \nonumber \\
&=&\Gamma_{\rm ren}(H;m,\mu)
-\Gamma_{\rm ren}(H;m=0,\mu) 
\end{eqnarray}
In (\ref{gammatilde}), from (\ref{constbar})
and (\ref{constren}),  
\begin{equation}
\Gamma_{{\rm ren}}(H;m=0,\mu) = {V H^2 \over 16 \pi^2}
\left(\frac{1}{6} \ln ({\mu^2 \over H}) - 2 \zeta'(-1)\right),
\end{equation}
where $\zeta'(s)$ is the first derivative of Riemann 
zeta function and $\zeta^\prime(-1)\approx -0.165421$. 
Notice that $Q(H;m)$ is 
independent of the normalization mass $\mu$ and is well-behaved in
the small mass limit. Explicitly, it is given by the expression
\begin{equation} \label{gammatilde2}
Q(H;m)= 2V\int^{m^2}_0 {d \bar{m}^2}
 \int^\infty_0 d s
 {e^{-\bar{m}^2 s} \over(4\pi s)^2}
\left[{(Hs/2)^2 \over \sinh^2(Hs/2)} -1 \right],
\end{equation}
and in the small mass limit, this leads to  
\begin{eqnarray}\label{gammarenzero}
Q(H;m)
&=&{VH^2\over 8 \pi^2} \int^{{m^2/H}}_0 d \bar{m}^2 
\left[-1/2-  \bar{m}^2 ( \log\bar{m}^2+ \gamma)+\cdots \right] 
\nonumber \\
&=& {VH^2 \over16 \pi^2}
\left[ -m^2/H -(m^2/H)^2(\log(m^2/H)-1/2+\gamma)
+\cdots 
\right].
\end{eqnarray}

In Fig.2, graphs for $Q(H;m)$, the exact one and those based on 
approximation schemes, are 
given as  functions of
$X \equiv m/\sqrt{H}$. The exact result, i.e., that based on the expression 
(\ref{gammatilde2}) is
represented by a solid line, which
exhibits a monotonically decreasing behavior starting from the maximum at
$X=0$. Clearly the small mass expansion up to ${\cal O}(m^4/H^2)$ provides 
a reliable approximation for $X \alt 0.4$, while the large mass expansion for  
$Q(H;m)$,
\begin{equation}
Q(H;m) = {VH^2 \over16 \pi^2}\left(-\frac{1}{6} \ln ({m^2\over H}) 
+2 \zeta'(-1) -{1\over 120} ({H\over m^2})^2
+ {1\over 504 } ({H\over m^2})^4
-{1\over 720} ({H\over m^2})^6
+\cdots \right),
\end{equation}
(this formula is obtained from (\ref{inversemass}), (\ref{constren}) 
and (\ref{gammatilde})) can be trusted in the range 
$X \agt 1$. In the intermediate region $0.4 \alt X \alt 1$ 
the large-mass expansion 
curve (a long dashed line in Fig.2) may then be smoothly
connected to that 
 given from the small-$m$ expansion (\ref{gammarenzero}),
assuming a monotonic behavior (as should be reasonable for a simple
background field). Evidently, with this interpolation, one could have
acquired a nice overall fit over the entire mass range even if the exact
curve were not known. We also see from Fig.2 the typical behaviors which 
are shown by the small-mass or large-mass expansion curves.
\begin{figure}
\begin{center}
\epsfxsize=4in
\epsffile{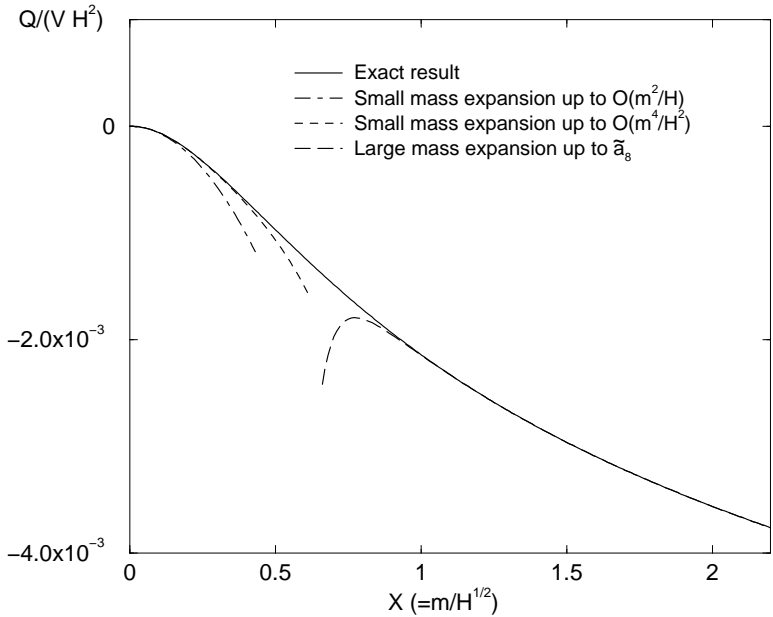}
\caption{Plot of  $Q(H;m)$. }
\end{center}
\end{figure}

\section{Large mass expansion for the spin-0 instanton effective action
~~~~~~~~~~~~~~~~~~~~~~~~~~~~~~~~~~~~~~~~~~~~~~~~~~~~~~~~~~~~~~~~~~~~~~
~~~~~~~~~~~~~}

We now turn to the case of a BPST instanton 
background\cite{belavin},
i.e., a self-dual solution of Yang-Mills field equations given by
\begin{equation}\label{instansol}
A_\mu(x)\equiv A^a_\mu(x){\tau^a\over 2}= { \eta_{\mu\nu a}\tau^a
x_\nu \over
x^2 +\rho^2},
\end{equation}
where $\eta_{\mu\nu a}\;(a=1,2,3)$ are the so-called 'tHooft
symbols\cite{thooft} and $\rho$ denotes the size of the instanton. 
The associated field strength  $F_{\mu\nu}$ is
\begin{equation} \label{fmunu}
F_{\mu\nu}=-2{\rho^2\eta_{\mu\nu a}\tau^a \over (x^2 +\rho^2)^2}.
\end{equation}
In this instanton background, the exact expression for the one-loop 
effective action due to a spin-0 or spin-1/2 matter field of nonzero 
mass is not known; only the result in the massless limit is known 
\cite{thooft}. This quantity will be studied with the help of 
approximation schemes in this paper. Specifically, taking the matter 
field to be that of an isospin-1/2, spin-0 particle, the corresponding 
effective action is studied using the large mass expansion in this 
section and by the small mass expansion in the next section. In 
Sec.{\rm VI}, we then use these results for a spin-0 matter field 
to obtain the corresponding results appropriate to a spin-1/2 matter 
field (i.e., quark). Note that, in the case of a spin-1/2 matter field, 
a {\it direct} application of the small mass expansion can 
be very subtle due to the presence of normalizable zero modes for 
the massless Dirac equation \cite{schwarz}). 

 The large mass expansion for the spin-0 effective action is described 
by our formula (\ref{gammabarmass}). To use this formula, one needs 
to know some higher-order coefficients in the series (\ref{series}), 
with 
$G^{-1} = -D^2$ and the instanton background given above. Calculations 
of these higher-order Seeley-DeWitt coefficients are straightforward 
in principle, but get very involved as the order increases. Fortunately, 
thanks to the rapidly growing computer capacity to handle a large 
number of terms in the symbolic calculations, the explicit expressions 
for the Seeley-DeWitt coefficients in general background fields have 
been found recently up to the sixth order \cite{belkov,fliegner,van}. We 
will utilize these results for our calculations below. 

 In the instanton background (\ref{instansol}) the renormalized 
one-loop effective action $\bar{\Gamma}(A)$, defined by (\ref{gammabar}), 
will be a function of $m\rho$ only. Hence our large mass expansion is 
really an expansion in $1/m^2\rho^2$. Also the expressions for the 
Seeley-DeWitt coefficients are simplified considerably if we take 
into account the fact that our background field satisfies the classical 
Yang-Mills equations of motion. For such on-shell background fields, 
the space-time integral of the Seeley-DeWitt coefficients $\tilde{ a}_n(x), 
n=3,4,5$ (for a spin-0 
matter field) are given as \cite{van}

\begin{eqnarray} \label{A3}
&&\int d^4 x \tilde{a}_3(x) = {i\over 90} \int d^4 x \  {\rm tr} 
\left[F_{\mu\nu} 
F_{\nu\rho} F_{\rho\mu}\right], 
\\\nonumber \\
\int d^4 x &~& \tilde{a}_4(x) =
{1\over 24} \int d^4 x \  {\rm tr}\left[ 
 {17\over 210} F_{\mu\nu} F_{\mu\nu} F_{\lambda\kappa} F_{\lambda\kappa}
 +{2\over 35} F_{\mu\nu} F_{\nu\rho} F_{\mu\lambda} F_{\lambda\rho} 
+{1\over105 } F_{\mu\nu} F_{\nu\rho} F_{\rho\sigma} F_{\sigma\mu}\right.
\nonumber \\ 
&~&\hspace{3 cm} \qquad \left.+{1\over 420} 
F_{\mu\nu} F_{\rho\sigma} F_{\mu\nu} 
F_{\rho\sigma}\right],
\label{A4} 
\\\nonumber\\
\int d^4 x &~& \tilde{a}_5(x) = {1\over 120}\int d^4 x \ {\rm tr}\left[ 
i{1\over 945}  F_{\mu\nu} F_{\rho\sigma} F_{\tau\mu} 
F_{\nu\rho} F_{\sigma\tau}
-i{47\over 126}  F_{\mu\nu} F_{\mu\nu} F_{\rho\sigma} 
F_{\sigma\tau} F_{\tau\rho}\right.
\nonumber \\
&~&+i{1\over 126}  F_{\mu\nu} F_{\rho\sigma} F_{\mu\nu} 
F_{\sigma\tau} F_{\tau\rho} 
+i{1\over 63}  F_{\mu\nu} F_{\nu\rho} F_{\mu\sigma} 
F_{\sigma\tau} F_{\tau\rho} 
-i{11\over 189} F_{\mu\nu} F_{\rho\sigma} F_{\sigma\nu} 
F_{\mu\tau} F_{\tau\rho}  
\nonumber \\
&~&+i{37\over 945}  F_{\mu\nu} F_{\nu\rho} F_{\rho\sigma} 
F_{\sigma\tau} F_{\tau\mu}
+{4\over 189} F_{\nu\tau} F_{\tau\sigma}
( D_\mu F_{\nu\rho}) (D_\mu F_{\rho\sigma}) 
 -{2\over 63} F_{\lambda\kappa} (D_\mu F_{\nu\rho}) F_{\nu\rho} (D_\mu 
F_{\lambda\kappa})  
\nonumber \\
&~&-{2\over 189}  F_{\lambda\sigma} (D_\mu F_{\nu\rho}) F_{\rho\sigma} (D_\mu 
F_{\nu\lambda})  
+{4\over 63}  F_{\sigma\tau} F_{\sigma\tau}
(D_\mu F_{\nu\rho}) (D_\mu F_{\nu\rho}) 
+{2\over 63} F_{\mu\tau} F_{\tau\sigma}
(D_\mu F_{\nu\rho}) (D_\sigma F_{\nu\rho}) 
\nonumber \\
&~&\left.+{4\over 189} F_{\sigma\tau} F_{\tau\nu}
( D_\mu F_{\nu\rho}) (D_\mu F_{\rho\sigma}) 
\right].
\label{A5}
\end{eqnarray}
Note that the on-shell expressions for the space-time integral of 
$ \tilde{a}_3(x)$  and $ \tilde{a}_4(x)$ involve only  
the field strength, while that for $ \tilde{a}_5(x)$ involves the 
derivatives of the field strength also. For the expression of 
$\tilde{a}_6(x)$, which occupies more than a page, 
see Ref.\cite{fliegner}. In Appendix A, the expressions valid without 
using the classical equations of motion (and before the space-time 
integration) can also be found.

Inserting the expression (\ref{fmunu}) for the field strength into 
the formulas (\ref{A3}) and (\ref{A4}) and carrying out  
tensor algebra and trace calculations, we find
\begin{eqnarray}
\int d^4 x \ \tilde{a}_3(x) &~& 
= \int d^4x \ {64 \rho^6 \over 15(x^2 + \rho^2)^6} 
={16 \over 75} {\pi^2\over \rho^2}, \\
\int d^4 x \ \tilde{a}_4(x)&~&
=\int d^4x \ {544 \rho^8\over 35 (x^2 + \rho^2)^8}
={272\over735 } {\pi^2\over \rho^4}.
\end{eqnarray}
The next coefficient $\tilde{a}_5$ involves the covariant derivative of field 
strength,
\begin{equation}\label{delfmunu}
D_\lambda F_{\mu\nu}={4\rho^2 \tau^a\over (x^2 + \rho^2)^3}
[2\eta_{\mu\nu a} x_\lambda-\eta_{\lambda \mu a}x_\nu
+\eta_{\lambda\nu a}x_\mu + \delta_{\lambda\mu}
\eta_{\nu\sigma a} x_\sigma
- \delta_{\lambda\nu} \eta_{\mu\sigma a} x_\sigma].
\end{equation}
Calculations
of higher-order Seeley-DeWitt coefficients with the instanton background
can be very laborious. Together with the formulas given
above and that in Ref.\cite{fliegner} for $\tilde{a}_6(x)$, we have thus 
used the
``Mathematica'' program to do the necessary trace calculations as well as
tensor algebra. From the expression for the $\tilde{a}_5$ coefficient, 
we obtain the result
\begin{eqnarray}
\int d^4x\,\tilde{a}_5(x)&=&\int d^4x\, 
{512 (35 x^2 \rho^8 - 39 \rho^{10})\over 315(x^2 + \rho^2)^{10}}
\nonumber \\
&=&-{1856\over2835} {\pi^2\over\rho^6},
\end{eqnarray}
while, for the $\tilde{a}_6(x)$ term, 
\begin{eqnarray}
\int d^4x\,\tilde{a}_6(x)
&=&-\int d^4x\, {256\over 51975(\rho^2 +x^2)^{12}} \nonumber \\
&& \times
[397710x^8\rho^4 -765270 x^6\rho^6 + 404961 x^4\rho^{8}
-86418x^2\rho^{10} +2876\rho^{12}]
\nonumber \\
&=&
{63328\over444675} {\pi^2\over\rho^{8}}.
\end{eqnarray}

Based on the explicit calculations given above, 
we obtain the following large-mass expansion for $\overline{\Gamma}(m\rho)$:
\begin{equation} \label{gammabarlargemass}
\overline{\Gamma}(m\rho)=-{1\over75}{1\over m^2\rho^2}
-{17\over735}{1\over m^4\rho^4}
+{232\over2835}{1\over m^6\rho^6}
-{7916\over148225}{1\over m^8\rho^8}
+\cdots
\end{equation}
\begin{figure}
\begin{center}
\epsfxsize=4in
\epsffile{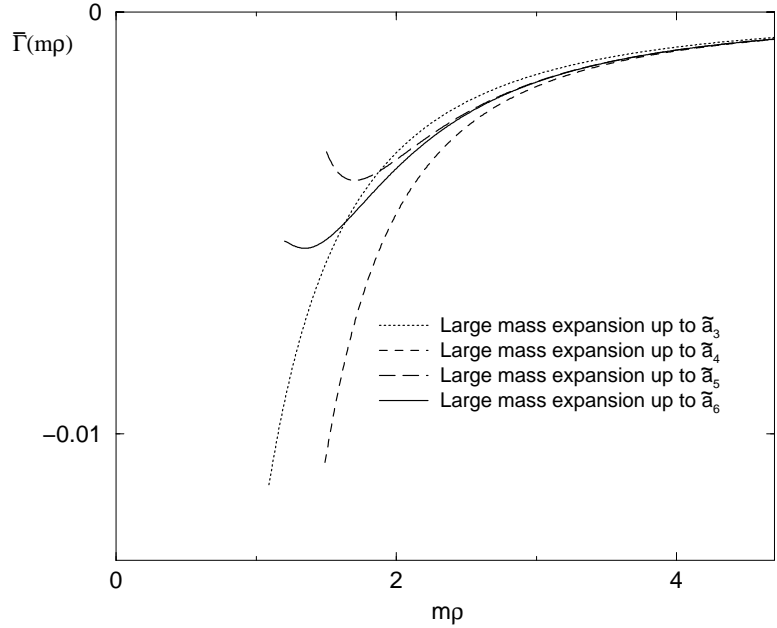}
\caption{Plot of $\overline{\Gamma}(m\rho)$ for the instanton background.}
\end{center}
\end{figure}
\begin{flushleft}In Fig.3 we have given the 
plots based on this expression (first keeping 
only the $\tilde{a}_3$-term, then including the $\tilde{a}_4$-term 
also, etc). This is a useful approximation when $m\rho$ is large, say, 
$m\rho \agt K$. What would be the lower-end value $K$ here ? In the 
absence of the exact expression for $\bar{\Gamma}(m\rho)$, a possible 
criterion for telling the validity range of the series 
(\ref{gammabarlargemass}) will be as follows. If $A_l$ denotes the 
${\cal O}({1\over (m\rho)^{2 l}})$ term in the series and 
$\Gamma_l\equiv \sum^l_{n=1}A_n$, we may demand that the series 
(\ref{gammabarlargemass}) remain {\it stable} in the sense that  
the relative importance of each newly added term decreases, i.e., 
$1=\Big|{A_1\over \Gamma_1}\Big| > \Big|{A_2\over \Gamma_2}\Big| > 
\Big|{A_3\over \Gamma_3}\Big| > \Big|{A_4\over \Gamma_4}\Big| > 
\cdots$.  As this criterion is used, we obtain the (conservative) 
value $K\simeq 1.8$. The result of large mass expansion may thus be 
trusted in the range given by $m\rho \agt 1.8$.

\end{flushleft}

\section{Small mass expansion for the spin-0 instanton effective action
~~~~~~~~~~~~~~~~~~~~~~~~~~~~~~~~~~~~~~~~~~~~~~~~~~~~~~~~~~~
~~~~~~~~~~~~~~~~~~~~~~~~~~}

 For small $m\rho$,\ that is, $m\rho$ significantly below $1$, the 
one-loop effective action in the instanton background (\ref{instansol}) 
can be studied with the help of the small mass expansion or mass 
perturbation, since its exact expression in the massless limit is 
known. Here we shall denote the corresponding spin-0 effective action 
$\Gamma_{{\rm ren}}$, which is defined by (\ref{gammaren}), as 
$\Gamma_{{\rm ren}}(m,\rho,\mu)$. For $m=0$ we have, from 
the computations of 
'tHooft \cite{thooft}, 
\begin{equation}  \label{Grenm0}
\Gamma_{{\rm ren}}(m=0,\rho,\mu) = {1\over 6}\ln \mu\rho + \alpha(1/2) 
\end{equation}
with $\alpha(1/2)={1\over 6}\gamma + {1\over6}\ln\pi - {1\over\pi^2}
\zeta^{\prime}(2)- {17\over72}\simeq 0.145873$. Our goal in this section is to 
compute explicitly the ${\cal O}(m^2)$ term of 
$\Gamma_{{\rm ren}}(m,\rho,\mu)$. Note that this small-$m$ approximation for 
$\Gamma_{{\rm ren}}(m,\rho,\mu)$ contains a non-analytic piece in $m$ and so 
it is not a trivial task to extract the desired term. 
 
 Our first task is to develop a small mass expansion for 
$\Gamma_{{\rm ren}}(m,\rho,\mu)$, which is finite at every order. For the 
purpose it is convenient to consider its derivative with respective 
to $m^2$, i.e., 
$\partial\Gamma_{{\rm ren}} / \partial m^2$, which is 
independent of the normalization mass $\mu$. The latter, 
being equal to the $m^2$-derivative of our regularized effective action 
$\Gamma(A)$, will have the proper-time representation 
\begin{equation} \label{derG}
{\partial\Gamma_{{\rm ren}}(m,\rho,\mu) \over \partial m^2} = 
\lim_{\eta \to 0^+}\int^{\infty}_{\eta} ds \ e^{-m^2s}\int d^4x 
 \ {\rm tr}\Big[\Big\langle xs\Big |x\Big > - 
\Big\langle xs\Big |x\Big >_{A_{\mu}=0}\Big],
\end{equation} 
where we have used (\ref{propertime}). Here note that 
$\Big\langle xs\Big |x\Big >\equiv \lim_{y \to x}\Big\langle xs\Big 
|y\Big >$ is nonsingular as long as 
$s > 0$. Paying due attention to various (singular) limits involved,
it is possible to recast the formula (\ref{derG}) into that 
involving ordinary spin-0 propagators 
\begin{equation} 
G(x,y;m^2)\equiv \Big\langle x \Big |{1\over -D^2+m^2}\Big |y\Big >, 
\hskip 1cm 
G_0(x,y;m^2)\equiv \Big\langle x \Big |{1\over -\partial^2+m^2}\Big |y\Big >.
\end{equation}
The explicit formula, which can be derived from (\ref{derG}),
reads\cite{lee82}
\begin{equation} \label{derGa0}
{\partial\Gamma_{{\rm ren}}(m,\rho,\mu) \over \partial m^2} = 
\int d^4x\lim_{y \to x} {\rm tr}[G(x,y;m^2) - a_0(x,y)G_0(x,y;m^2)],
\end{equation}
where $a_0(x,y)$ is the zeroth order coefficient in the asymptotic series
(\ref{series}). For small $(x-y)_\mu,\ a_0(x,y)$ has the following expression
\begin{eqnarray} \label{a0}
a_0(x,y) = &~&I + i(x-y)_\mu A_\mu(y) + {i\over 4}(x-y)_\mu(x-y)_\nu [
\partial_\mu A_\nu(y)+\partial_\nu A_\mu(y) + i\{A_\mu(y),A_\nu(y)\}]
\nonumber \\
&~& +{\cal O}((x-y)^3).
\end{eqnarray}
Presence of the $a_0G_0$ term in (\ref{derGa0}) guarantees a 
finite result for $\partial\Gamma_{{\rm ren}}(m,\rho,\mu) / \partial m^2$.

 For small $m$, one may then try to evaluate the right-hand-side of 
(\ref{derGa0}) by exploiting the appropriate expansion of the propagators 
in $m^2$ and the known exact massless propagation function in the 
instanton background. We shall denote the latter by 
$\bar{G}(x,y)\equiv\Big\langle x \Big |{1 \over -D^2}\Big |y\Big >$. But a naive expansion of the 
form 
\begin{eqnarray} 
G(x,y;m^2) =&~&\bar{G}(x,y)-m^2\int d^4z \bar{G}(x,z)
\bar{G}(z,y) + m^4\int d^4z d^4w \bar{G}(x,z)
\bar{G}(z,w)\bar{G}(w,y) \nonumber \\&~&+\cdots
\end{eqnarray}
is not valid since, aside from the leading term, all other terms of 
this series involve logarithmically divergent integrals. (Note that 
$\bar{G}(x,z) = {\cal O}({1 \over |z|^2}$) as $|z| \to \infty$). Moreover,
the $m^2=0$ limit of $\partial \Gamma_{{\rm ren}}/ \partial m^2$ does 
not exist in the instanton background since, according to explicit 
calculations (see the comment after (\ref{logdiv}) below), the integral 
in the right hand side of (\ref{derGa0}) for $m^2=0$ diverges 
logarithmically. This indicates that, as $m^2 \to 0$, 
$\Gamma_{{\rm ren}}(m,\rho,\mu)$ approaches the 'tHooft result 
(\ref{Grenm0}) in a non-analytic manner. To resolve this problem, 
we shall below describe an alternative expansion scheme (which utilizes 
the idea of Carlitz and Creamer\cite{carlitz} in a suitable form).

 The expansion we shall use has the form 
\begin{equation} \label{propa}
{1\over -D^2+m^2} = {1\over (-D^2)}(-\partial^2){1\over -\partial^2+m^2} 
\left[\sum^\infty_{r=0}\left(-m^2\left\{{1\over (-D^2)}-
{1\over (-\partial^2)}\right\}(-\partial^2)
{1\over -\partial^2+m^2}\right)^r ~\right].
\end{equation}
This can be derived in the following way. First observe that  
\begin{equation} \label{expanpropa} 
{1\over -D^2+m^2} = {1\over-\partial^2+m^2}+ (-D^2){1\over -D^2+m^2} 
\left\{{1\over(-D^2)}-{1\over(-\partial^2)}\right\} (-\partial^2) 
{1\over-\partial^2+m^2}. 
\end{equation}
Then, using the identity 
\begin{equation}
(-D^2){1\over-D^2+m^2} = 1- m^2 {1\over-D^2+m^2},
\end{equation}
it is not difficult to see that (\ref{expanpropa}) can be rewritten as 
\begin{equation}
{1\over -D^2+m^2} = {1\over(-D^2)}(-\partial^2){1\over-\partial^2+m^2} - 
m^2{1\over-D^2+m^2}\left\{{1\over(-D^2)}- {1\over(-\partial^2)}
\right\}(-\partial^2) {1\over-\partial^2+m^2}.
\end{equation} This last equation may be solved for ${1\over-D^2+m^2}$ 
in an iterative manner, and the result is 
the expansion (\ref{propa}). Evidently, (\ref{propa}) is an expansion 
in powers of $m^2\{{1\over(-D^2)}-{1\over(-\partial^2)}\}(-\partial^2)
{1\over-\partial^2+m^2}$, and we expect that this will yield a 
convergent series for small $m$ if the background field is such that 
\begin{equation}
\Big\langle x \Big |\left\{{1\over(-D^2)}-
{1\over(-\partial^2)}\right\}(-\partial^2)\Big |y\Big > 
\to 0~({\rm sufficiently \ fast}), ~{\rm as}\ |x-y| \to \infty.
\end{equation}
In the case of the instanton background, this means that we have to 
work with the expression given in the singular gauge, i.e.,
\begin{equation} \label{singular}
A_\mu(x) = {\rho^2\bar{\eta}_{\mu\nu a}\tau_a x_\nu\over
x^2(x^2+\rho^2)}.
\end{equation}
[Here, $\bar{\eta}_{\mu\nu a}$ differs from $\eta_{\mu\nu a}$ only by the 
sign in the components with $\mu$ or $\nu$ equal to 4] This is allowed by 
the gauge invariance of the effective action.

 If only the leading term of (\ref{propa}) is used in (\ref{derGa0}), 
we now find that 
\begin{eqnarray} \label{derG3}
{\partial\Gamma_{{\rm ren}}(m,\rho,\mu) \over \partial m^2}\nonumber =&~&
 \int d^4x \lim_{y \to x} {\rm tr}\left[\Big\langle x \Big 
|{1\over(-D^2)}(-\partial^2)
{1\over-\partial^2+m^2}\Big |y\Big > \right.\\&~&\left.
- a_0(x,y)\Big\langle x \Big |{1\over-\partial^2+m^2}\Big |y\Big >
\right] + {\cal O}(m^2).
\end{eqnarray}
This is not yet in the convenient form for actual computations. So, 
based on the following observations 
\begin{eqnarray}
&~&\Big\langle x \Big |{1\over(-D^2)}(-\partial^2){1\over-\partial^2+m^2}\Big |y\Big >
= \Big\langle x \Big |{1\over(-D^2)}\Big |y\Big > + \Big\langle x \Big |{1\over(-D^2)}(-\partial^2)\left\{
{1\over-\partial^2+m^2}-{1\over(-\partial^2)}\right\}\Big |y\Big >,\nonumber \\
&~&\\
&~&a_0(x,y)\Big\langle x \Big |{1\over-\partial^2+m^2}\Big |y\Big > 
= a_0(x,y)\Big\langle x \Big | {1\over(-\partial^2)}\Big |y\Big > + 
\Big\langle x \Big |\left\{{1\over-\partial^2+m^2} -
{1\over(-\partial^2)}\right\}\Big |y\Big > \nonumber \\   
&~&~~~~~~~~~~~~~~~~~~~~~~~~~~~~~~~~~~~~
+ ({\rm terms \ vanishing} \ {\rm as}\ y \to x) ,
\end{eqnarray}
we make suitable rearrangements in the right hand side of (\ref{derG3})
to obtain the following formula (to be used for computations)
\begin{equation} \label{derG4}
{\partial\Gamma_{{\rm ren}}(m,\rho,\mu) \over \partial m^2}\nonumber 
 = \int d^4x \left[\lim_{y \to x} {\rm tr}\left\{\Big\langle x 
\Big |{1\over(-D^2)}\Big |y\Big >
- a_0(x,y)\Big\langle x \Big |{1\over(-\partial^2)}\Big |y\Big >\right\} -J(x)\right]  + {\cal O}(m^2),
\end{equation}
where the function $J(x)$ is given by 
\begin{equation} \label{J1}
J(x) = -\int d^4z \  {\rm tr}\left\{\Big\langle x 
\Big |\left({1\over(-D^2)} - {1\over(-\partial^2)}
\right) \Big |z\Big >(-\overleftarrow{\partial_z}^2)\Big\langle z \Big |\left({1\over-\partial^2+m^2}
- {1\over(-\partial^2)}\right)\Big |x\Big >\right\} .
\end{equation}
But for the $J(x)$ term, what we have in the right hand side of (\ref{derG4})
is just the (logarithmically divergent) expression representing 
${\partial\Gamma_{{\rm ren}}(m,\rho,\mu) \over \partial m^2}|_{m^2=0}$ (see 
(\ref{derGa0})). As we shall see below, this divergence is tamed by the 
additional term $J(x)$. The very structure of $J(x)$ given in (\ref{J1})
also ensures that it is free of any short-distance divergence. 

 The first term inside the integrand of (\ref{derG4}) is evaluated as 
follows. The spin-0 (and isospin-1/2) massless propagator in the instanton 
background (\ref{singular}) is given by \cite{brown}
\begin{equation}\label{G1} 
\bar{G}(x,y) \equiv \Big\langle x \Big |{1\over-D^2}\Big |y\Big > = {1\over 4\pi^2 (x-y)^2}
 {1+{\rho^2(x\cdot y+i\bar{\eta}_{\mu\nu a}x_\mu y_\nu\tau_a) \over
 x^2y^2} \over \sqrt{1+{\rho^2\over x^2}}\sqrt{1+{\rho^2\over y^2}}}
\end{equation}
Then, writing $x=y+\epsilon$, we find after some straightforward 
calculations 
\begin{eqnarray} \label{masslessp} 
\Big\langle x \Big |{1\over-D^2}\Big |y\Big > &=& {1\over 4\pi^2\epsilon^2} \left[ 
\left(1+{\rho^4(y\cdot\epsilon)^2 -\rho^2y^2(y^2+\rho^2)\ep^2\over
2(y^2)^2(y^2+\rho^2)^2}\right)\right. \nonumber \\ 
&~&~~~~~~+\left.\left(1-{(2y^2+\rho^2)(y\cdot\ep)
\over y^2(y^2+\rho^2)}\right) {i\rho^2\bar{\eta}_{\mu\nu a}\tau_a 
\ep_\mu y_\nu \over y^2(y^2+\rho^2)}\right]
+ {\cal O}(\epsilon).
\end{eqnarray}
On the other hand, if (\ref{singular}) is inserted into the expression 
(\ref{a0}), we have 
\begin{eqnarray} \label{a02} 
a_0(x,y) &=& I + i(x-y)_\mu{\rho^2\bar{\eta}_{\mu\nu a}\tau_a y_\nu 
\over y^2(y^2+\rho^2)} \nonumber \\
 &~&-{1\over 2}(x-y)_\mu(x-y)_\nu 
{\rho^4(y^2\delta_{\mu\nu}-y_\mu y_\nu) + 2i\rho^2(2y^2+\rho^2)
\bar{\eta}_{\mu\lambda a} \tau_a y_\nu y_\lambda \over
(y^2)^2(y^2+\rho^2)^2} + {\cal O}((x-y)^3)
\end{eqnarray}
and therefore 
\begin{eqnarray}\label{tra0}  
{\rm tr}\left\{a_0(x,y)  \Big\langle x \Big |
{1\over(-\partial^2)}\Big |y\Big >\right\} = 
{1\over 2\pi^2\ep^2} \left(1-{\rho^4 y^2\ep^2-\rho^4(y\cdot\ep)^2 
\over 2(y^2)^2(y^2+\rho^2)} \right) +{\cal O}(\epsilon) 
\end{eqnarray}
From (\ref{masslessp}) and (\ref{tra0}), we thus obtain the following 
expression 
\begin{equation} \label{logdiv}
\lim_{y \to x} {\rm tr}\left\{\Big\langle x \Big |{1\over(-D^2)}\Big |y\Big >
- a_0(x,y)\Big\langle x \Big |{1\over(-\partial^2)}\Big |y\Big >\right\} = -{\rho^2\over 4\pi^2 
(x^2 +\rho^2)^2}. 
\end{equation}
[We here remark that the result (\ref{logdiv}) is unchanged even if one 
takes the regular-gauge instanton solution (\ref{instansol}) as 
the background field]. Clearly, with this term alone, the remaining 
$x$-integration would yield a logarithmically divergent result.  

 We now turn to the evaluation of $J(x)$. Noting that 
\begin{equation}\label{momzx}
\Big <z\Big | \left({1\over -\partial^2 + m^2} - {1\over (-\partial^2)}
\right) \Big|x \Big> = - m^2\int {d^4 p\over (2\pi)^4} 
{e^{-i p\cdot (z-x)}\over p^2 (p^2 +m^2)},
\end{equation}
(\ref{J1}) may be rewritten as 
\begin{equation}\label{J2}
J(x) = m^2 \int {d^4 p\over (2\pi)^4} 
{e^{i p\cdot x}\over p^2 (p^2 +m^2)} F(x,p)
\end{equation}
with $F(x,p)$ given by 
\begin{eqnarray}\label{Fxp}
F(x,p) &~& = \int d^4 z e^{-i p\cdot z} {\rm tr} \left\{ \Big<x\Big| 
\left({1\over (-D^2)} - {1\over (-\partial^2)}\right)\Big|z \Big> 
(-\overleftarrow{\partial_z}^2)\right\} \nonumber \\
&~& = 2 \int d^4z \  e^{-i p\cdot z} (-\overrightarrow{\partial_z}^2) 
\left[{1\over 4 \pi^2 (x-z)^2} \left( {1 + 
{\rho^2 x\cdot z\over x^2 z^2}\over \sqrt{1+\rho^2/x^2} 
\sqrt{1+\rho^2/z^2}} - 1 \right)\right].
\end{eqnarray}
In (\ref{Fxp}) we have used  the expression (\ref{G1}) and the 
factor 2 at front arose from the isospin trace. We are here interested in
${\cal O}(1)$ or ${\cal O}(\log m^2)$ contribution to the right hand side of 
(\ref{derG4}). Let us see when and where such contribution can arise, 
based on our formulas (\ref{J2}) and (\ref{Fxp}). For any 
finite $x$-value, the function $F(x,p)$ is well-behaved for 
all $p$. Due to the overall multiplicative factor $m^2$ in 
(\ref{J2}), $J(x)$ for finite $x$ (or, more precisely, for $x^\mu$ 
satisfying the condition $|x| \ll {1\over m}$) would then 
be ${\cal O}(m^2)$ and hence 
no desired contribution.  It is thus sufficient 
to study $J(x)$ for 
large $x$, i.e., $x^\mu$ in the region $|x| > L$ with $\rho \ll 
L \ll m^{-1}$. Now, due to the factor $e^{i p\cdot x}/p^2(p^2+m^2)$ within the 
integrand of (\ref{J2}), we further conclude that the small-$p$ region of
$F(x,p)$, with $|x| >L$, can be the source for the desired 
contribution; if the contribution from the region $|p| \alt m$ is 
excluded from the right hand side of (\ref{J2}), $J(x)$ becomes 
${\cal O}(m^2)$.

 To study the function $F(x,p)$ for $|x| > L$ (with $\rho \ll L \ll m^{-1}$)
and $|p| \alt m$, we write $F(x,p)$ as the sum of its value at $p=0$ 
plus the correction term, viz.,
\begin{equation} \label{FC}
F(x,p) = F(x,p=0) + C(x,p)
\end{equation}
Then, from (\ref{Fxp}),
\begin{mathletters}\begin{eqnarray}
F(x,p=0) &=& 2 \int d^4z \  (-\overrightarrow{\partial_z}^2) 
\left[{1\over 4 \pi^2 (x-z)^2} \left( {1 + 
{\rho^2 x\cdot z\over x^2 z^2}\over \sqrt{1+\rho^2/x^2} 
\sqrt{1+\rho^2/z^2}} - 1 \right)\right] \label{Fx01}\\
&=& - \lim_{R \to \infty} {1\over 2 \pi^2} \oint_{|z|=R} d^3 \Omega R^3 
{z_\mu \over R} \left[ {2(x-z)_\mu\over [(x-z)^2]^2} 
\left({1 + {\rho^2 x\cdot z\over x^2 z^2}\over \sqrt{1+\rho^2/x^2}
\sqrt{1+\rho^2/z^2}} -1\right) \right.\nonumber \\
&~& \hspace{4 cm} 
 +\left.{1\over (x-z)^2} {\partial \over \partial z_\mu} 
\left( {1 + {\rho^2 x\cdot z\over x^2 z^2}\over \sqrt{1+\rho^2/x^2}
\sqrt{1+\rho^2/z^2}} \right) \right],\label{Fx02} 
\end{eqnarray} \end{mathletters}
\begin{flushleft}where we used Gauss's law. [Note that, for
 very large $|z|$, the integrand 
in (\ref{Fx01}) behaves like ${\cal O}({1\over |z|^5})]$. Evaluating the 
surface integral in (\ref{Fx02}) immediately gives \end{flushleft}
\begin{equation} 
F(x,p=0) = 2 \left({1\over \sqrt{1+\rho^2/x^2}}-1 \right),
\end{equation}
and hence 
\begin{equation}
F(x,p=0) = -{\rho^2\over x^2} + {\cal O} ({\rho^4\over x^4}), 
\qquad {\rm for} \ |x| > L.
\end{equation}
Now, inserting the thus evaluated $F(x,p=0)$ for $F(x,p)$ in
(\ref{J2}), we obtain the following contribution to $J(x)$: 
\begin{eqnarray}\label{J3}
J(x) &=& \theta(|x|-L) \int d^4 p \left({1\over p^2} - {1\over p^2+m^2}
\right) e^{i p\cdot x} \left(-{\rho^2\over x^2}\right) \ + \ {\cal O}(m^2) 
\nonumber \\
&=& -{\rho^2\over 4 \pi^2 (x^2)^2}\left[1-m|x| K_1 (m|x|)\right] 
\theta (|x| -L) \ + \ {\cal O}(m^2).
\end{eqnarray}
(Note that we have assumed $mL \ll 1$). On the other hand, it is 
possible to show (see Appendix B) that $C(x,p)$ in (\ref{FC}) is at most 
${\cal O}(|p|L{\rho^2\over x^2})$ or ${\cal O}({\rho^4\over x^3 L})$ or
${\cal O}({\rho^4\over x^4})$, when $|x| >L$ and $|p| \alt m$. With 
this finding used in (\ref{J2}), it is easy to see that no ${\cal O}(1)$ or 
${\cal O}(\log m)$ contribution results from the $C(x,p)$ part of 
$F(x,p)$. Thus, to the order we want, our formula (\ref{J3}) has no 
further correction. 

 Evidently, if the contribution in (\ref{J3}) is considered together 
with that in (\ref{logdiv}), the $x$-integration in (\ref{derG4}) 
will give a finite result. Furthermore, since the function $F(x,p)$ does
not involve mass $m$ at all, the scale $L$ we introduced 
can be chosen, for $m\rho \to 0$, such that 
$\rho\ll  L\ll  1/m$. With this understanding, we may now perform the 
integral in the right hand side of (\ref{derG4}) to secure the 
unambiguous result 
\begin{equation}\label{derG5}
{\partial \Gamma_{{\rm ren}}(m,\rho,\mu) \over \partial m^2} = 
{\rho^2 \over 2} \ln (m\rho) + {\rho^2 \over 2} (\gamma + {1\over 2} - \ln2)
 + {\cal O}(m^2 \rho^4).
\end{equation}
Then, based on this formula and the 'tHooft result (\ref{Grenm0}),
we immediately obtain the desired small-mass expansion for 
$\Gamma_{{\rm ren}}(m,\rho,\mu)$:
\begin{eqnarray}\label{smallG}
\Gamma_{{\rm ren}}(m,\rho,\mu) &~&= \Gamma_{{\rm ren}}(m=0,\rho,\mu)
+ \int^{m^2}_0 d\bar{m}^2 {\partial \Gamma_{{\rm ren}}(\bar{m},\rho,\mu)
\over \partial \bar{m}^2} \nonumber \\ 
&~& = {1\over 6} \ln (\mu\rho) + \alpha(1/2) + {(m\rho)^2 \over 2}
\left[ \ln(m\rho) + \gamma - \ln 2\right] + {\cal O}((m\rho)^4).
\end{eqnarray}
The ${\cal O}((m\rho)^2 \ln(m\rho))$ term in this formula was first 
found in Ref.\cite{carlitz}, while the ${\cal O}((m\rho)^2)$ term 
without the $\ln (m\rho)$ factor is new.

\section{Mass interpolation and the spin-1/2 instanton effective action
~~~~~~~~~~~~~~~~~~~~~~~~~~~~~~~~~~~~~~~~~~~~~~~~~~~~~~~~~~~~~~~~~~
~~~~~~~~~~~~~~~~~}
In the previous two sections the spin-0 instanton effective action were 
computed for relatively large $m\rho$ and for small $m\rho$. The result 
can be summarized by 
\begin{equation}\label{Gren}
\Gamma_{{\rm ren}}(m,\rho,\mu) = \frac{1}{6}\ln (\mu\rho) 
+ \alpha(1/2) + Q(m\rho)
\end{equation}
with the quantity $Q$, a function of $m\rho(\equiv X)$ only, 
behaving as
\begin{equation}\label{smallandlarge}
Q(X) =
\left \{ \begin{array} {ll} {1\over2}X^2 \ln X + {1\over2}
(\gamma - \ln 2)
X^2 + \cdots, &\qquad (X \alt 0.5) \\
-{1\over6} \ln X - \alpha({1\over2}) - \frac{1}{75} {1\over X^2} 
-\frac{17}{735} {1\over X^4} + \frac{232}{2835} {1\over X^6} 
-\frac{7916}{148225}{1\over X^8} + \cdots,
&\qquad (X \agt1.8). \end{array} \right.
\end{equation}
(Note that, in the instanton background, (\ref{gammaren}) implies that 
$\Gamma_{{\rm ren}}(m,\rho,\mu) = \frac{1}{6} \ln (\mu/m) + 
\bar{\Gamma}(m\rho)$). In the indicated validity ranges of $X$, 
the function $Q(X)$ is plotted in Fig.4. We have here assumed that our 
small mass expansion in (\ref{smallG}) can be used reliably for 
$X \alt 0.5$; this estimate is based on measuring the effect of 
the $(m\rho)^4$ term (with the numerical coefficient taken to be 
${\cal O}(1)$) against the terms which appear explicitly 
in (\ref{smallG}). 
\begin{figure}
\begin{center}
\epsfxsize=4in
\epsffile{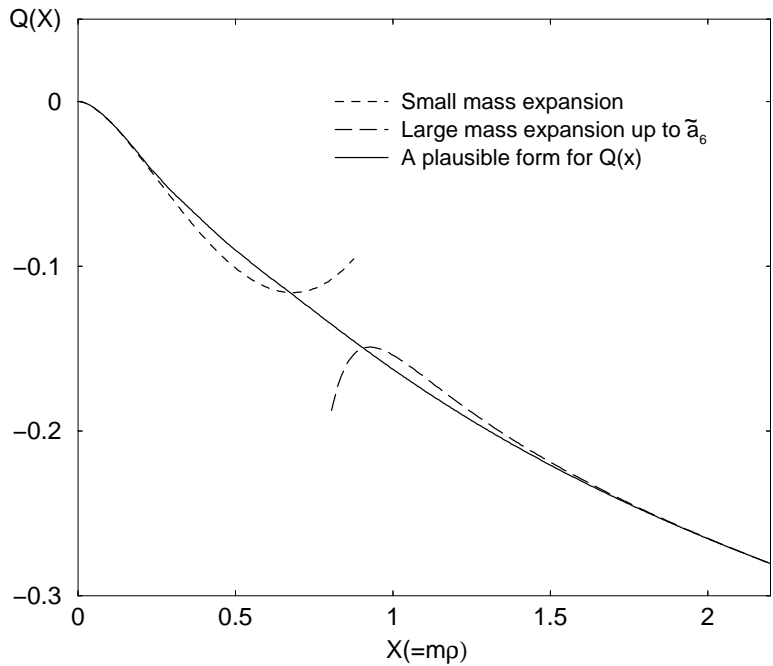}
\caption{Plot of  $Q(m\rho)$. }
\end{center}
\end{figure}

 Now, what could be said on the behavior of the function $Q(X)$ in 
the intermediate region $0.5 \alt X \alt 1.8$? Since the background 
field under consideration has a smooth profile, one naturally 
expects that $Q(X)$ also be a smooth function of $X$; that is, $Q(X)$ 
would be represented by a smooth interpolating curve connecting the 
known forms of the curve in the regions $X \agt 1.8$ and $X\alt 0.5$. Let 
us further assume that the region for interpolation, 
$0.5 \alt X \alt 1.8$, can be viewed as being reasonably small. Then, 
looking at how $Q(X)$ actually behaves for $X \agt 1.8$ and 
$X \alt 0.5$ (see Fig.4), it appears to be quite plausible to 
suppose that $Q(X)$ is a {\it monotonically decreasing} function 
of $X$ for all $X >0$. But, since we have in no way proved this 
monotonic behavior in the presence of the instanton 
background, one may regard this as a conjecture.[For instance, 
the possibility that $Q(X)$ may develop a local maximum or 
minimum within the range $ 0.5 \alt X \alt 1.8$ is not 
excluded. Incidentally, such monotonic behavior was also observed 
in the case of a self-dual 
constant field strength (see Fig.2)]. Accepting the conjecture, it 
might be useful (especially for phenomenological analysis of 
instanton effects) to have a certain smooth function $Q(X)$ in the 
entire range $X >0$ which meets this requirement. With the plausible
curve for $Q(X)$ taken by that given in Fig.4, we have found (after some 
trial and errors) that it may be described by the function of the 
form  
\begin{equation} \label{Qfit}
Q(X) \sim  -\frac{1}{6} \ln X - \alpha + 
{\frac{1}{6}\ln X + \alpha - (3 \ \alpha + \beta) X^2 
- \frac{1}{5} X^4\over 1 - 3 X^2 + 20 X^4 + 15 X^6}, 
\qquad (\mbox{for all} \ X>0)
\end{equation}
with $\alpha \equiv \alpha(1/2) \simeq 0.145873$ and 
$\beta = \frac{1}{2} (\ln 2 - \gamma) \simeq 0.05797$. This form 
incorporates correctly the small-$X$ and leading large-$X$ 
behaviors shown in (\ref{smallandlarge}). For the 
tunneling 
amplitude which is more directly related to $e^{-\Gamma_{{\rm ren}}}$, 
this amounts to using the expression (\ref{Qfit}) with
\begin{equation} \label{expGren}
e^{-\Gamma_{{\rm ren}}(m,\rho,\mu)} = (\mu\rho)^{-1/6} e^{-\alpha(\frac{1}{2})
- Q(m\rho)} \qquad \qquad (\mbox{for arbitrary mass} \ m).
\end{equation}

 Various results obtained for the spin-0 field case can be used to 
derive the corresponding results appropriate to the spin-1/2 
one-loop instanton effective action. The latter will be needed if 
one wishes to 
consider the loop correction to the vacuum tunneling amplitude in
QCD due to quark fields. In a self-dual Yang-Mills background, the 
hidden supersymmetry of the system allows one to express the 
spin-1/2 proper-time Green function $\langle x s|y\rangle^{(1/2)} \equiv 
\langle x| e^{- s(\gamma D)^2} |y\rangle$ in terms of the corresponding 
spin-0 function 
$\langle x s|y\rangle^{(0)} \equiv \langle x| e^{- s(- D^2)} 
|y\rangle$ (with 
the same isospin representation assumed). Explicitly, this is 
described by the operator relation \cite{lee82} 
\begin{equation}\label{spinor}
e^{-s(\gamma D)^2} = e^{-s(-D^2)} {1 + \gamma_5\over 2} + \gamma D 
{1\over - D^2} e^{-s(-D^2)} \gamma D {1 - \gamma_5\over 2} + P,
\end{equation}
where $P$ is the projection operator into the zero mode subspace 
of $\gamma D$ and can be expressed by $P = (1- \gamma D {1\over - D^2} 
\gamma D)({1 - \gamma_5\over 2})$\cite{brown}. Using 
the relation (\ref{spinor}) with the definition of $\Gamma_{{\rm ren}} 
(A)$ (see Sec.II), it is then possible to derive a simple relationship 
between the spin-1/2 and spin-0 one-loop effective actions. If
$\Gamma_{{\rm ren}}^{(1/2)}(A) \ (\Gamma_{{\rm ren}}^{(0)}(A)) $
denotes the one-loop effective action as defined by (\ref{gammaren}) 
for a spin-1/2 (complex spin-0) field of mass $m$ in a self-dual 
Yang-Mills background, we have in fact 
\begin{equation}
\Gamma_{{\rm ren}}^{(1/2)}(A) = -\frac{1}{2} n_F \ln ({m^2\over \mu^2}) 
- 2 \Gamma_{{\rm ren}}^{(0)}(A) 
\end{equation}
or, for the respective contributions to the tunneling amplitude 
\begin{equation}
e^{-\Gamma_{{\rm ren}}^{(1/2)}(A)} = ({m\over \mu})^{n_F} 
e^{2 \Gamma_{{\rm ren}}^{(0)}(A)},
\end{equation}
where $n_F$ is the number of normalizable spinor zero modes in 
the given background\cite{schwarz}. Now, using the result (\ref{Gren}) 
for $\Gamma_{{\rm ren}}^{(0)}(A)$, we have the spin-1/2 instanton 
effective action expressed as (with $n_F =1$)  
\begin{equation}\label{spinorG}
\Gamma_{{\rm ren}}^{(1/2)}(m,\rho,\mu) = - \ln {m \over \mu} - 
\frac{1}{3} \ln \mu\rho - 2 \alpha(1/2) - 2 Q(m\rho),
\end{equation}
or, for the tunneling amplitude,
\begin{equation}\label{expspinorG}
e^{-\Gamma_{{\rm ren}}^{(1/2)}(m,\rho,\mu)} = {m\over\mu}
(\mu \rho)^{1/3} e^{2 \alpha(1/2) + 2 Q(m \rho)}, 
\end{equation}
where $Q(m\rho)$ is the function specified in 
(\ref{smallandlarge}) (and represented in Fig.4).

 The expression in (\ref{spinorG}) or (\ref{expspinorG}) describes the 
one-loop contribution to the vacuum tunneling by an isospin-1/2 
quark field of mass $m$. If one accepts our conjecture, the function $Q(m\rho)$ may be taken as a monotonically decreasing function of 
$m\rho$ which has the limiting behaviors as given in 
(\ref{smallandlarge}). The 
renormalization prescription appropriate to the expression (\ref{spinorG}) 
is that specified by (\ref{gammaren}). If one wishes to obtain the 
corresponding amplitude in the minimal subtraction in the dimensional regularization scheme, the finite renormalization counterterm (see (\ref{spinconnect}))
\begin{equation}
{1\over (4\pi)^2\cdot 6} (\ln 4 \pi -\gamma)\int d^4 x F^a_{\mu\nu} 
F^a_{\mu\nu}  = \frac{1}{3}(\ln 4 \pi - \gamma)
\end{equation}
must be added to the expression (\ref{spinorG}). Thus, in the minimal 
subtraction scheme, the amplitude due to a spin-1/2 quark of mass $m$ 
reads 
\begin{equation}
e^{-\Gamma_{{\rm MS}}^{(1/2)}(m,\rho,\mu)}
 = {m\over \mu}\left( {\mu\rho\over 4 \pi e^{-\gamma}}\right)^{1/3}
e^{ 2\alpha(1/2) + 2 Q(m\rho)}.
\end{equation}
With $Q(m\rho)$ set to zero, this reduces to the result of 
'tHooft\cite{thooft}.  For applications to the 
real QCD with the SU(3) gauge group, one must 
also take into account the well-known group theoretical factor associated 
with various ways of embedding the SU(2) instanton solution 
\cite{bernard}.

\section{Discussions}
 In this work we studied the massive quark contribution to the one-loop 
instanton effective action in QCD. For this purpose, we made use of the 
approximation scheme valid for relatively large mass as well as the small-mass expansion. These considerations provide a reliable approximation to the 
one-loop effective action if the magnitude of $m\rho$ is such that 
$m\rho \agt 1.8$ or $m\rho \alt 0.5$. The expression for the effective 
action contains a function $Q(m\rho)$, the magnitude of which is 
uncertain in the range $0.5 \alt m\rho \alt 1.8$. Based on
the known behaviors of $Q(m\rho)$ in the ranges $m\rho \alt 1.8$ and 
$m\rho \alt 0.5$, we suggested that $Q(m\rho)$ be a smooth, monotonically 
decreasing function of $m\rho$. If the latter turns out to be true, a 
simple interpolation formula for $Q(m\rho)$ (as we considered in 
(\ref{Qfit})) suffice for considerations in most phenomenological 
analyses. 

What can be done to reduce the uncertainty in the function
 $Q(m\rho)$ for 
$0.5 \alt m\rho \alt 1.8$ ? With the explicit calculation of the 
${\cal O}((m\rho)^4)$ term in the small mass expansion, it should be 
possible to push the lower 
end of the uncertain range to a slightly higher value. On the 
other hand, we expect that including the next higher Seeley-DeWitt 
coefficient in the large -mass expansion would not bring a 
significant new information. More useful direction might be to try a direct 
numerical evaluation of the functional determinant, with the 
help of the scattering theory in a radially symmetric background 
field.(Some related techniques are discussed 
in Ref.\cite{bordag}) Perhaps, by some 
mathematical argument, it might also be possible to actually prove that  
the function $Q(m\rho)$, which 
is equal to $\Gamma_{{\rm ren}}(m,\rho,\mu) - \Gamma_{{\rm ren}} 
(m=0,\rho,\mu)$ ( for a spin-0 field) in an instanton background, is 
a monotonically 
decreasing function of $m\rho$. These are left for further study.

\vskip .4in
{\bf Acknowldgments}
\vglue .2in

One of us(C.L.) wishes to thank the Korea Institute for Advanced Study for 
hospitality while part of this work was undertaken. This work was 
supported in part by the BK 21 project of the Ministry of
Education, Korea(O.K. and C.L.), by Korea Research Foundation 
Grant(C.L.), and by the research supporting program of the 
University of Seoul(H.M.).

\newpage
\appendix
\section{}

In the literature\cite{belkov,fliegner,van}, the Seeley-DeWitt 
coefficients $\tilde{a}_n(x)$ for 
a spin-0 or spin-1/2 matter field have been calculated up to $n=6$.
Here, for the case of a (complex) spin-0 field, we shall give the 
explicit expressions for $\tilde{a}_n(x)$ up to $n=5$ in a general 
off-shell background field. They read  
\begin{eqnarray}
\tilde{a}_3(x)&~& =-{1\over 6} \ {\rm tr} \left[ 
 i \frac{2}{15}  F_{\kappa\lambda}F_{\lambda\mu}F_{\mu\kappa}
-\frac{1}{20} (D_\kappa F_{\lambda\mu}) (D_\kappa F_{\lambda\mu}) 
\right],
\\ 
\tilde{a}_4(x)&~&   
 ={1\over 24} \ {\rm tr}\left[- \frac{1}{21} F_{\kappa\lambda}F_{\lambda\mu}
F_{\mu\nu}F_{\nu\kappa} 
+  \frac{11}{420} F_{\kappa\lambda}F_{\mu\nu}F_{\lambda\kappa} F_{\nu\mu}
 +  \frac{2}{35} F_{\kappa\lambda}F_{\lambda\kappa}F_{\mu\nu}F_{\nu\mu} 
\right.\nonumber \\ 
&~&+  \frac{4}{35} F_{\kappa\lambda}F_{\lambda\mu}F_{\kappa\nu}F_{\nu\mu} 
 + i \frac{6}{35}  F_{\kappa\lambda}(D_\mu F_{\lambda\nu})(D_\mu F_{\nu\kappa}) 
 + i \frac{8}{105}  F_{\kappa\lambda}(D_\lambda F_{\mu\nu})
(D_\kappa F_{\nu\mu}) 
\nonumber \\
 &~&\left. 
 +  \frac{1}{70} (D_\kappa D_\lambda F_{\mu\nu})(D_\lambda D_\kappa F_{\nu\mu}) 
\right], 
\\
\tilde{a}_5(x)&~& =  
-{1\over 120} \ {\rm tr}\left[
 -  i \frac{2}{945}  F_{\kappa\lambda}F_{\lambda\mu}
F_{\mu\nu}F_{\nu\rho}F_{\rho\kappa} 
-  i\frac{8}{63}  F_{\kappa\lambda}F_{\lambda\mu}F_{\kappa\nu}
F_{\mu\rho}F_{\rho\nu}\right.
\nonumber \\
&~&+ i \frac{16}{945}  F_{\kappa\lambda}F_{\mu\nu}F_{\lambda\rho}
F_{\nu\kappa}F_{\rho\mu} 
+ i \frac{22}{189}  F_{\kappa\lambda}F_{\lambda\mu}F_{\kappa\nu}
F_{\nu\rho}F_{\rho\mu} 
 + i \frac{31}{378}  F_{\kappa\lambda}F_{\lambda\mu}
F_{\nu\rho}F_{\mu\kappa}F_{\rho\nu} 
\nonumber \\
&~& + i \frac{53}{378}  F_{\kappa\lambda}F_{\lambda\kappa}
F_{\mu\nu}F_{\nu\rho}F_{\rho\mu} 
+\frac{1}{9} F_{\kappa\lambda}F_{\lambda\mu}
(D_\nu F_{\kappa\rho})(D_\nu F_{\rho\mu})
 + i \frac{1}{18}  F_{\kappa\lambda}(D_\lambda D_\mu F_{\nu\rho})
(D_\mu D_\kappa F_{\rho\nu})
\nonumber \\
&~& - \frac{1}{189} F_{\kappa\lambda}F_{\lambda\mu}
(D_\kappa F_{\nu\rho})(D_\mu F_{\rho\nu}) 
+ \frac{1}{252} (D_\kappa D_\lambda D_\mu F_{\nu\rho})
(D_\mu D_\lambda D_\kappa F_{\rho\nu}) 
 +  \frac{1}{378} F_{\mu\kappa}F_{\rho\nu}F_{\kappa\lambda}
(D_\lambda D_\mu F_{\nu\rho}) 
\nonumber \\
 &~&+ i \frac{2}{21}  (D_\kappa F_{\lambda\mu})
(D_\kappa D_\nu F_{\mu\rho})(D_\nu F_{\rho\lambda}) 
+  \frac{2}{63} F_{\kappa\lambda}(D_\mu F_{\nu\rho})
F_{\lambda\kappa}(D_\mu F_{\rho\nu}) 
+ i \frac{4}{63}  F_{\kappa\lambda}(D_\mu D_\nu F_{\lambda\rho})
(D_\nu D_\mu F_{\rho\kappa}) 
\nonumber \\
&~& - \frac{5}{63} F_{\kappa\lambda}F_{\lambda\mu}(D_\nu 
F_{\mu\rho})(D_\nu F_{\rho\kappa}) 
 +  \frac{5}{63} F_{\kappa\lambda}(D_\mu F_{\lambda\nu})
F_{\kappa\rho}(D_\mu F_{\rho\nu}) 
+  \frac{5}{63} F_{\kappa\lambda}(D_\mu F_{\nu\rho})
F_{\rho\lambda}(D_\mu F_{\nu\kappa}) 
\nonumber \\
&~&+i \frac{5}{126}  (D_\lambda F_{\rho\nu}) (D_\kappa F_{\lambda\mu})
(D_\mu D_\kappa F_{\nu\rho})  
-\frac{10}{189} F_{\kappa\lambda}(D_\lambda F_{\mu\nu})
F_{\nu\rho}(D_\rho F_{\mu\kappa}) 
- \frac{8}{189} F_{\rho\mu} F_{\kappa\lambda}(D_\lambda F_{\mu\nu})
(D_\nu F_{\kappa\rho}) 
\nonumber \\
&~& + i \frac{5}{126}  (D_\kappa F_{\lambda\mu})(D_\mu F_{\nu\rho})
(D_\lambda D_\kappa F_{\rho\nu}) 
- \frac{10}{189} F_{\kappa\lambda}(D_\mu F_{\lambda\nu})
F_{\mu\rho}(D_\kappa F_{\rho\nu}) 
+  \frac{11}{189} F_{\kappa\lambda}(D_\mu F_{\lambda\kappa})
F_{\nu\rho}(D_\mu F_{\rho\nu}) 
\nonumber \\
&~&+\frac{11}{189} F_{\kappa\lambda}(D_\mu F_{\nu\rho})
F_{\rho\nu}(D_\mu F_{\lambda\kappa}) 
- \frac{11}{378} F_{\rho\nu} F_{\kappa\lambda}F_{\lambda\mu}
(D_\mu D_\kappa F_{\nu\rho}) 
+  \frac{13}{252} F_{\kappa\lambda}F_{\lambda\kappa}
(D_\mu F_{\nu\rho})(D_\mu F_{\rho\nu}) 
\nonumber \\
&~&-\frac{16}{63} F_{\kappa\lambda}(D_\lambda F_{\mu\nu})
F_{\nu\rho} (D_\kappa F_{\rho\mu}) 
- \frac{16}{189} F_{\kappa\lambda}F_{\lambda\mu}
F_{\nu\rho}(D_\mu D_\kappa F_{\rho\nu}) 
 - \frac{19}{756} F_{\kappa\lambda}(D_\lambda F_{\mu\nu})
F_{\kappa\rho}(D_\rho F_{\nu\mu}) 
\nonumber \\
&~&-\frac{19}{756} F_{\kappa\lambda}(D_\mu F_{\nu\rho})
F_{\mu\lambda}(D_\kappa F_{\rho\nu}) 
 +  \frac{25}{189} F_{\rho\nu} F_{\kappa\lambda}(D_\mu F_{\lambda\nu})
(D_\mu F_{\kappa\rho}) 
- \frac{26}{189} F_{\rho\nu} F_{\kappa\lambda}(D_\mu F_{\lambda\nu})
(D_\rho F_{\mu\kappa}) 
\nonumber \\
&~&-\frac{34}{189} F_{\rho\mu} F_{\kappa\lambda}(D_\lambda F_{\mu\nu})
(D_\rho F_{\nu\kappa}) 
- \frac{41}{378} F_{\kappa\lambda}F_{\lambda\mu}(D_\mu 
F_{\nu\rho})(D_\kappa F_{\rho\nu}) 
 +  \frac{61}{756} F_{\rho\nu} F_{\kappa\lambda}(D_\mu F_{\lambda\kappa})
(D_\mu F_{\nu\rho}) 
\nonumber \\
&~&\left.+  \frac{61}{756} F_{\kappa\lambda}F_{\mu\nu}(D_\rho 
F_{\lambda\kappa})(D_\rho F_{\nu\mu})
\right], 
\end{eqnarray}
where 
$D_\lambda F_{\mu\nu} \equiv [D_\lambda, F_{\mu\nu}]$
and $ D_\kappa D_\lambda 
F_{\mu\nu} \equiv [D_\kappa,[D_\lambda, F_{\mu\nu}]], \ \ {\rm etc.}$.

\section{}

The function $C(x,p)$ in (\ref{FC}) is given by
\begin{equation} \label{appC}
C(x,p) = 2 \int d^4z \left(e^{-i p\cdot z} - 1\right) 
\left(-\overrightarrow{\partial_z}^2\right)\left[{1\over 4 \pi^2(x-z)^2} 
\left({1 + {\rho^2 x\cdot z\over x^2 z^2}\over 
\sqrt{1+\rho^2/x^2}\sqrt{1+\rho^2/z^2}} -1\right)\right],
\end{equation}
and we are here interested in its behavior for $|x|>L$(with 
$\rho \ll L\ll m^{-1})$ and $p \alt m$. We divide this quantity into 
two parts, i.e., $C(x,p) = C_<(x,p) + C_>(x,p)$, where $C_<(x,p)$ 
denotes the contribution with the region of integration restricted to 
$|z| \leq L_1$(with $\rho \ll L_1 <L$) and $C_>(x,p)$ that from 
the region $|z|>L_1$.(We take $L_1$ to be of the same order 
as $L$.) Then, 
for $C_<(x,p)$, it will be safe to make an approximation 
$e^{- i p\cdot z} -1 \simeq -i p \cdot z$ (i.e., ${\cal O}(|p|L)$ at most)
inside the integrand of ({\ref{appC}) and so we find immediately 
\begin{equation}
C_<(x,p) = {\cal O}(|p|L {\rho^2\over x^2}), \qquad ({\rm for} \ |x| > L).
\end{equation}    
On the other hand, for $C_>(x,p)$, we may expand the factor 
${1\over \sqrt{1+\rho^2/x^2}\sqrt{1+\rho^2/z^2}}$ in the integrand of 
(\ref{appC}) as a power series in $\rho/|x|$ and $\rho/|z|$, and then 
\begin{eqnarray}
&~&{1\over 4 \pi^2(x-z)^2} 
\left({1 + {\rho^2 x\cdot z\over x^2 z^2}\over 
\sqrt{1+\rho^2/x^2}\sqrt{1+\rho^2/z^2}} -1\right)
\nonumber \\
&~&\longrightarrow  
- {\rho^2\over 8 \pi^2 x^2 z^2} + {\rho^4\over 
4 \pi^2(x-z)^2} {3(x^2)^2 + 3(z^2)^2 + 2x^2z^2 
-4 x\cdot z(x^2+z^2) \over 8 (x^2)^2(z^2)^2} + \cdots .\label{Cexpand}
\end{eqnarray}
As the differential operator $-\overrightarrow{\partial_z}^2$ acts on 
this expression, the first term in the right hand side of (\ref{Cexpand}) 
can be dropped. Thus, we may write 
\begin{eqnarray}
C_>(x,p) &=& 2 \int_{|z| > L_1} d^4 z \left(e^{-i p\cdot z} - 1\right) 
\nonumber \\
&~&\cdot \left(-\overrightarrow{\partial_z}^2\right)
\left[{\rho^4\over 4 \pi^2(x-z)^2} 
{3(x^2)^2 + 3(z^2)^2 + 2x^2z^2 
-4 x\cdot z(x^2+z^2) \over 8 (x^2)^2(z^2)^2} + \cdots \right],
\end{eqnarray}
and, for $|x|>L$ and $|p| \alt m$, it is not difficult to show 
that this can only lead to terms of ${\cal O}(|p|L 
{\rho^4\over x^2 L^2})$ or  
${\cal O}( {\rho^4 \over x^3 L})$ or 
${\cal O}( {\rho^4 \over x^4 })$. Hence, $C(x,p)$ is at 
most ${\cal O} 
(|p| L {\rho^2\over x^2})$ or ${\cal O}({\rho^4\over x^3 L})$ or 
${\cal O}({\rho^4\over x^4 })$.

%%%%%%%%%%%%%%%%%%%%%%%%%%%%%%%%%%%%%%%%%%%%%%%%%%%%%%%%%%%%%%%%%%%%%%%%%%

\end{document}